\documentclass{elsart}

\usepackage{natbib}

\usepackage{amssymb,epsfig}

\begin{document}

\begin{frontmatter}



\title{On the tuning of a wave-energy driven oscillating-water-column seawater pump to polychromatic waves}


\author[icmyl,pmmh]{Ramiro Godoy-Diana}\ead{ramiro@pmmh.espci.fr}
\author[icmyl]{  Steven P.R. Czitrom\corauthref{cor1}}\ead{czitrom@mar.icmyl.unam.mx}
\corauth[cor1]{Corresponding author}

\address[icmyl]{Instituto de Ciencias del Mar y Limnolog\'\i a (ICMyL)\\ Universidad Nacional
Aut\'onoma de M\'exico (UNAM)\\ Ciudad Universitaria, 04510 M\'exico D.F., M\'exico}
\address[pmmh]{Physique et M\'ecanique des Milieux H\'et\'erog\`enes (PMMH)\\ UMR 7636 CNRS; ESPCI; Univ. Paris6; Univ. Paris7\\ 10 rue Vauquelin, F-75231 Paris Cedex 05, France}
\begin{abstract}
Performance of wave-energy devices of the oscillating water column (OWC) type is greatly enhanced when a resonant condition with the forcing waves is maintained. The natural frequency of such systems can in general be tuned to resonate with a given wave forcing frequency. In this paper we address the tuning of an OWC sea-water pump to polychromatic waves. We report results of wave tank experiments, which
were conducted with a scale model of the pump. Also, a numerical solution for the pump equations, which were
proven in previous work to successfully describe its behavior when driven by
monochromatic waves, is tested with various polychromatic wave spectra. Results of the numerical
model forced by the wave trains measured in the wave tank experiments are used to develop a tuning criterion for the sea-water pump.
\end{abstract}


\end{frontmatter}

\section{Introduction}

Conversion of ocean wave energy into a useful form has been studied intensely during the past two-decades, particularly regarding generation of electricity \citep[e.g.][]{ross1995,clement2002}. Special attention has been given to one class of wave energy converters based on an oscillating water column (OWC) \citep[see][for a review of the OWC physics]{falnes2002}. In such systems, seawater is forced by incident waves to oscillate inside a partially-submerged chamber or duct. The free surface within this chamber compresses a volume of air which is, in most cases, directed to flow through a turbine thus driving an electric generator \citep[e.g.][]{falcao1996,heath2000}. Most R \& D efforts on the subject have been focused on the wave energy absorption efficiency optimisation \citep[e.g.][]{evans1981,korde1999,perdigao2003}.

\begin{figure}[ht]
\begin{center}
  \epsfig{file=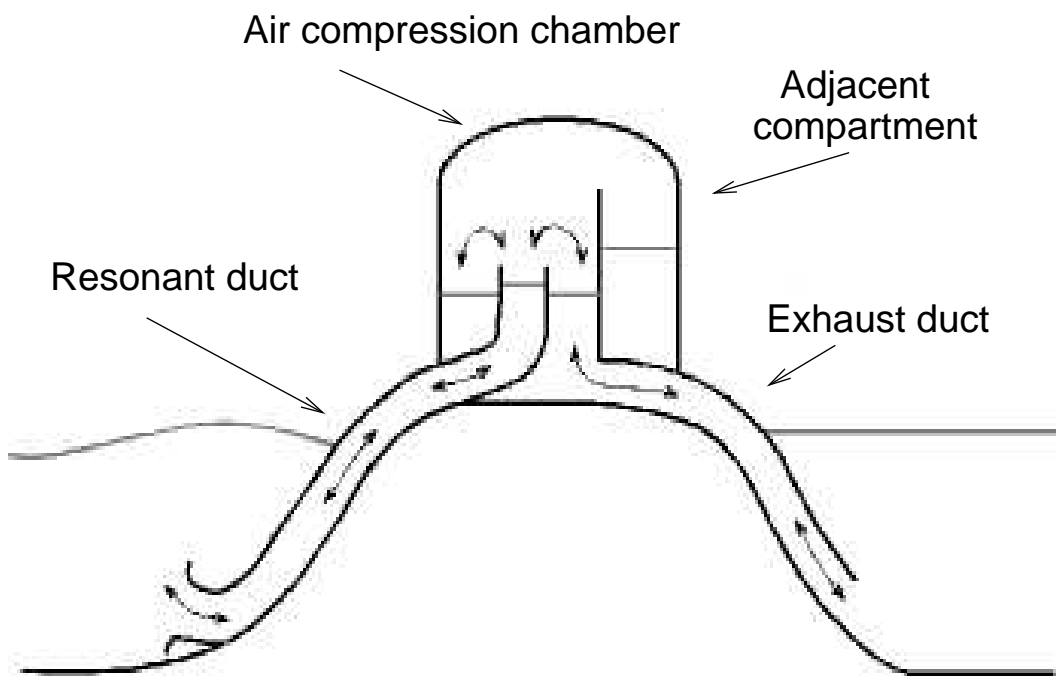,width=1\linewidth}
\end{center}
\caption{Schematic diagram of the seawater pump \citep[from][]{czitrom2000b}}
\label{figsibeo}
\end{figure}

An OWC wave energy device intended for sea-water
pumping, involving no generation of electricity, has been described by \cite
{czitrom2000a,czitrom2000b}. The system has potential for various coastal
management purposes, such as aquaculture, flushing out of contaminated areas
or the recovery of isolated coastal lagoons as breeding grounds. The sea-water pump (shown in figure \ref{figsibeo}) is composed of: a
resonant duct, in which the wave oscillation, compared to the wave motion, is magnified and rectified by overtopping of the water column; an
exhaust duct, through which water flow resulting from wave rectification is
channeled to a receiving water body; and a variable volume air compression
chamber coupling the two ducts. The wave-induced pressure at the
mouth of the resonant duct drives an oscillating flow that spills water into
the compression chamber, and thence through the exhaust duct, with each
passing wave. Air in the chamber behaves like a spring against which water
in the resonant and exhaust ducts oscillates. Maximum efficiency is attained
at resonance when the system natural frequency of oscillation coincides with
the frequency of the driving waves \citep[see e.g.][]{lighthill1979we}. A
resonant condition can be obtained for different wave frequencies by means
of a variable volume compression chamber that adjusts the hardness of the
air spring. Response of the system to monochromatic waves has been previously studied by \cite{czitrom2000b}. They developed a tuning
algorithm that predicts the optimal volume of air for the
compression chamber, for given values of wave period and amplitude and tidal
elevation. In this paper we address the problem of tuning the wave pump when forced by polychromatic waves. We report on wave tank experiments with a scale model of the pump as well as on the results of numerical simulations driven by real wave tank data. A tuning criterion is
derived to optimize the system performance in these conditions.

This paper is organized as follows: in section \ref{background} we recall the equations describing the seawater pump as well as previous results on the tuning to monochromatic waves. Results of the experiments with a scale model and polychromatic forcing are reported in section \ref{experiment}. Tuning to polychromatic waves using experimental data and a numerical model of the pump is discussed in section \ref{tuning} followed by concluding remarks in section \ref{concl}.

\section{Background}
\label{background}

\subsection{System Equations}

The hydrodynamic theory for the OWC seawater pump studied here is based on what has been referred to as a heaving rigid-piston model \citep{evans1981,brendmo1996,falnes2002}, where the free surface inside the resonant duct is assumed to behave as a massless rigid piston. When the incoming waves are long with respect to the interior cross-section of the resonant duct, the deformation of the internal free-surface is negligible and the OWC is well described by the heave velocity of the `rigid-piston'. The prediction of the system behavior is thus equivalent to that of more general models \citep[e.g.][]{evans1982,sarmento1985} that take into account the deformation of the internal free-surface. The pumping system was shown by \cite{czitrom2000a} to be accurately described by equations (\ref{syseqs1}-\ref{syseqs2}) for the evolution of the water surface elevation inside each duct, when there is no spilling in the compression chamber. These equations were derived
applying the time-dependent form of Bernoulli's equation to streamlines that
join the free surfaces inside the resonant and exhaust ducts with the
surfaces of the forcing and receiving water bodies, respectively \cite[see e.g.][chapter 4]{falnes2002}. Terms were added to account for viscous losses due to friction \cite[see also][]{perez1996}
and vortex formation \cite[see also][]{knott_mackley1980}, which are not considered in Bernoulli's equation, and
radiation damping \cite[see also][]{knott_flower1980}. At the time of pumping, fluid surges from the resonant
duct and spills into the exhaust side of the compression chamber. The fluid
surface above the resonant duct bulges upward inducing a back pressure on
the fluid in the duct, proportional to the bulge height. This effect has to be
taken into consideration when solving the equations throughout the full
cycle \citep{czitrom2000a}.

\begin{equation}
\hspace{-5mm}\mathcal{L}_1(\chi_1)\frac{d^2\chi _1}{d%
t^2}+\frac{A_{f1}}2\left( \frac{d\chi _1}{dt}\right)
^2+c_{D1}\frac{d\chi _1}{dt}%
\left| \frac{d\chi _1}{dt}\right| + C(\chi_1,\chi_2)+\frac{g\chi _1}{\mathcal{A}\Omega ^2} = \mathcal{W}(t) \; \label{syseqs1}
\end{equation}
\begin{equation}
\hspace{-5mm}\mathcal{L}_2(\chi_2)\frac{d^2\chi _2}{dt^2}+\frac{A_{f2}}2\left( \frac{d\chi
_2}{dt}\right) ^2+c_{D2}\frac{%
d\chi _2}{dt}\left| \frac{d\chi _2}{dt}\right| + C(\chi_1,\chi_2) +\frac {g\chi _2}{\mathcal{A}\Omega ^2}  =  0 \; \label{syseqs2}
\end{equation}

\noindent where

\begin{equation}
\mathcal{L}_1(\chi_1)=\left( \frac{L_1+T_d}{\mathcal{A}}+A_{f1}\chi _1\right) \;\;\;,\;\;\;
\mathcal{L}_2(\chi_2)=\left( \frac{L_2\frac{A_{c}}{A_D}+L_{c}}{\mathcal{A}}+A_{f2}\chi_2\right)
\label{ductlengths}
\end{equation}

\begin{equation}
C(\chi_1,\chi_2)=\frac{P_A-\rho gH}{\rho \mathcal{A}^2A_{f1}\Omega ^2}\left[ \left( 1-%
\mathcal{A}\frac{A_1A_{f1}\chi _1+A_{c}A_{f2}\chi _2}{V_0}\right)
^{-\gamma }-1\right] \; \label{chamber}
\end{equation}

Equations (\ref{syseqs1}-\ref{syseqs2}) are written in non-dimensional form by scaling
length and time with the amplitude $\mathcal{A}$ and frequency $\Omega$ of an average forcing wave so that the forcing in equation (\ref{syseqs1}) reads

\begin{equation}
\mathcal{W}(t)=\frac{g}{\mathcal{A}A_{f1}\Omega ^2 }\sin\left( \Omega t\right) \;.
\label{forcing}
\end{equation}

\noindent A magnifying
factor $A_f$ for the oscillation amplitude in each duct, with respect to the
forcing wave, is considered. Subscripts 1, 2 and $c$ in equations (\ref{syseqs1}-\ref{syseqs2}) correspond to the resonant and exhaust ducts and compression
chamber respectively. $\chi $ is the surface displacement, in either duct,
relative to its' equilibrium position in the compression chamber, and $L$
and $A$ are lengths and areas respectively. $L_1$ and $L_2,$ the resonant
and exhaust duct lengths, correspond to the physical length of the ducts
plus an added length due to edge effects at the mouths \citep[see][]{czitrom2000b}. In addition, $P_A$ is the atmospheric pressure,
$\rho $ is the sea-water density,
$g$ is the gravitational acceleration,
$\gamma=C_p/C_v = 1.4$ is the ratio of the specific heat capacities of compressed air,
$T_d$ is the height of sea level above the receiving body of water (including tides), $V_0$ is the compression chamber volume, $H$ is the height of the equilibrium position of the free surface in the compression chamber above the receiving
body of water and $c_{D1,2}$ is a coefficient of  nonlinear losses including vortex formation, friction and radiation damping. $L_1$
and $T_d$ are referred to the equilibrium position on the
exhaust side of the compression chamber.

\begin{figure}[t]
\begin{center}
  \epsfig{file=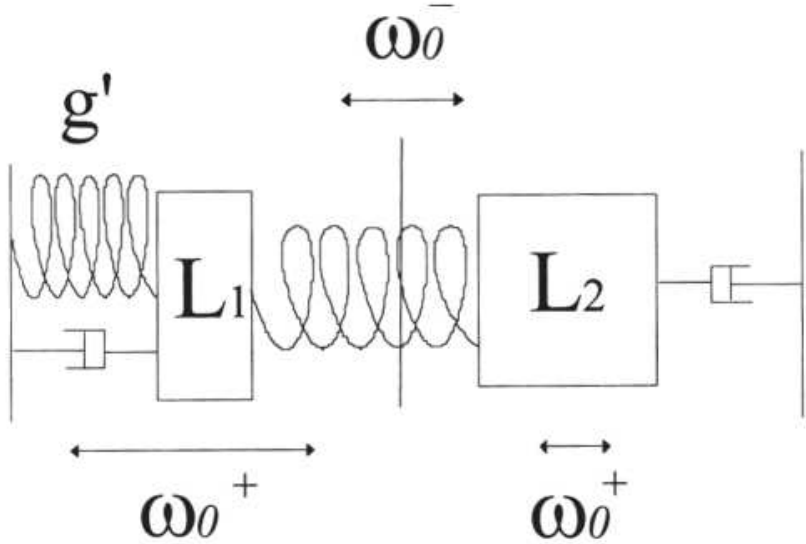,width=0.8\linewidth}
\end{center}
\caption{Representation of the pump as a damped two-mass two-spring oscillator \citep[from][]{czitrom2000b}}
\label{figresortes}
\end{figure}

The inertial term is first in both equations (\ref{syseqs1}-\ref{syseqs2}), where the water
masses in the ducts are written in terms of the duct lengths ---equation (\ref{ductlengths}). The second and
fifth terms come directly from Bernoulli's equation whereas the third term is added to represent the nonlinear losses due to vortex formation at the duct mouths, friction and radiation damping. Coupling of the two equations occurs through the air compression term (fourth in both
equations) and the wave forcing is included on the right hand side of the resonant duct equation (\ref{syseqs1}). A dimensional analysis of the pump equations performed by \cite{czitrom2000a} for a full-scale oceanic application, shows that the non-linear terms in equations (\ref{syseqs1}-\ref{syseqs2}) are
relatively small, and that their main contribution is to limit flow through
the pump. They showed that the pump behaves basically like a two-mass two-spring linear
system (figure \ref{figresortes}), where the two masses correspond to the
water in the two ducts and the springs represent gravity and the air
compression chamber. The restoring force on the oscillating system is thus provided by the compression of the air chamber and by the gravitational force. Non-linear losses, albeit small, are represented by the pistons.

\subsection{Numerical model}

\begin{figure}[t]
\begin{center}
  \epsfig{file=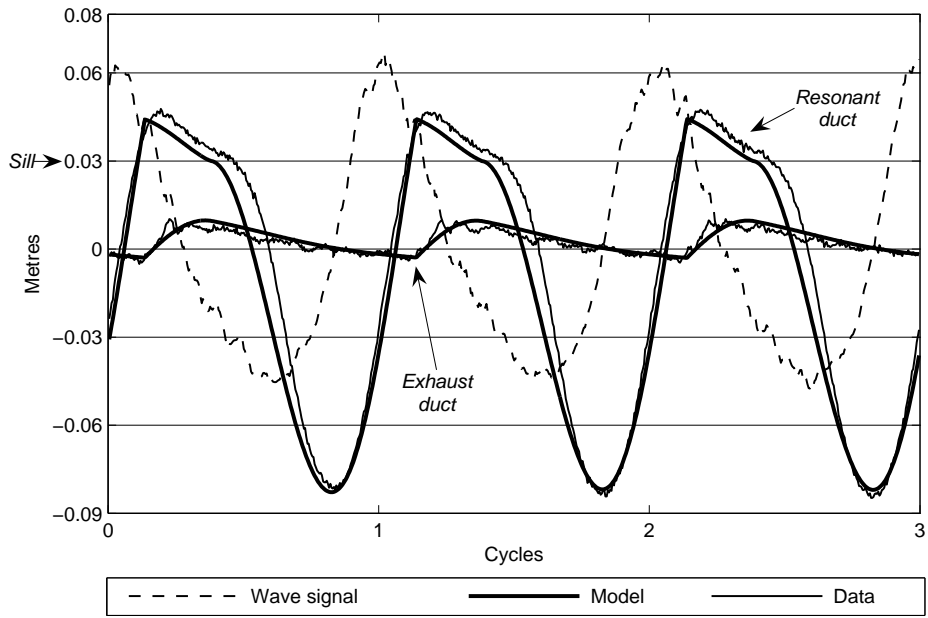,width=1\linewidth}
\end{center}
\caption{Modelled and observed resonant and exhaust duct water heights at the compression chamber and driving wave height plotted against time \citep[from][]{czitrom2000b}}
\label{figforcing}
\end{figure}

A numerical model based on a finite-difference integration of the pump equations (\ref{syseqs1}-\ref{syseqs2}), was thouroughly tested by \cite{czitrom2000b} for monochromatic forcing waves. They compared output from numerical model runs for the free surface position in the ducts and flow through the pump, with experimental results obtained in wave tank experiments with a physical model of the pump. They used the pressure signal measured at the duct mouth as forcing instead of the sinusoidal wave of equation (\ref{forcing}) and the system behaviour was very well predicted (see figure \ref{figforcing}).

\subsection{Tuning to Monochromatic Waves}

An expression for the natural frequencies of oscillation of the sea-water pump when there is
no spilling in the compression chamber can be derived from the linearized
version of equations (\ref{syseqs1}-\ref{syseqs2}). It reads \citep{czitrom2000b}:

\begin{equation}
\omega _0^{\pm }=\sqrt{\frac{\mathrm{A}\pm \sqrt{\mathrm{A}^2-4\mathrm{B}}}2}%
\;.  \label{frec}
\end{equation}

\noindent where,

\begin{equation}
\mathrm{A} =\frac{g+\alpha A_1}{\Omega ^2\left( L_1+T_d\right) }+\frac{%
\alpha A_{cam}}{\Omega ^2L_2}\;,
\end{equation}
\begin{equation}
\mathrm{B} =\frac{g\alpha A_{cam}}{\left( L_1+T_d\right) L_2\Omega ^4}\;,
\end{equation}
\begin{equation}
\alpha =\frac{\left( P_A-\rho gH\right) \gamma }{\rho V_0}\;.
\end{equation}

The two frequencies in equation (\ref{frec}) correspond to different modes
of oscillation of the water masses in the ducts. The higher frequency $%
\omega _0^{+}$, is associated to a synchronous oscillation of the two free
surfaces that compresses and decompresses the air chamber, and is the
natural frequency excited for pumping. The second mode, with the lower
frequency $\omega _0^{-}$, corresponds to a bodily oscillation of the two
masses about their center of mass.

Since expression (\ref{frec}) was derived from non-dimensional equations,
the resonant condition useful for pumping occurs at $\omega _0^{+}=1,$ and
the forcing wave frequency appears as $\Omega $ in coefficients A and B.
Solving for $V_0$ provides the following simple analytical algorithm for
tuning the system for resonance at the given wave frequency \citep{czitrom2000a}:

\begin{equation}
V_{0Lin}=\alpha \left[ \frac{A_r}{L_1\Omega ^2-g}+\frac{A_{cam}}{L_2\Omega ^2%
}\right] \;.  \label{volin}
\end{equation}

This tuning algorithm, however, is not completely adequate when there is
spilling in the compression chamber and the system is fully operational.
During pumping, the fluid that spills from the resonant duct ceases to weigh
on the rest of the fluid in that duct, so that the gravitational spring
restoring force becomes constant. This is equivalent to a
softening of the combined air-cushion-gravity restoring force, since
springiness is now only provided by the compression chamber. Pumping thus
decreases the natural frequency of oscillation of the system so that, in
order to maintain a resonant condition at the driving wave frequency, the
compression chamber spring must be hardened, by decreasing its' volume.
That is, resonance occurs at a volume of air smaller than the
linear prediction $V_{0Lin}$ when there is pumping.

The numerical model was used by \cite{czitrom2000b} to obtain a tuning algorithm for monochromatic
waves for the fully operational system. This algorithm
renders an optimal volume of air for the compression chamber ($V_0$), in
terms of the wave period and amplitude and the tidal elevation. The resonant air chamber volume ($V_0$) was thus found to increase with increasing
wave period and decreasing tidal height. They also showed that the flow rate through the pump at resonance ($Q_0$) increases with increasing wave period and wave amplitude.

\section{Experiments}
\label{experiment}

\begin{figure}[t]
\begin{center}
  \epsfig{file=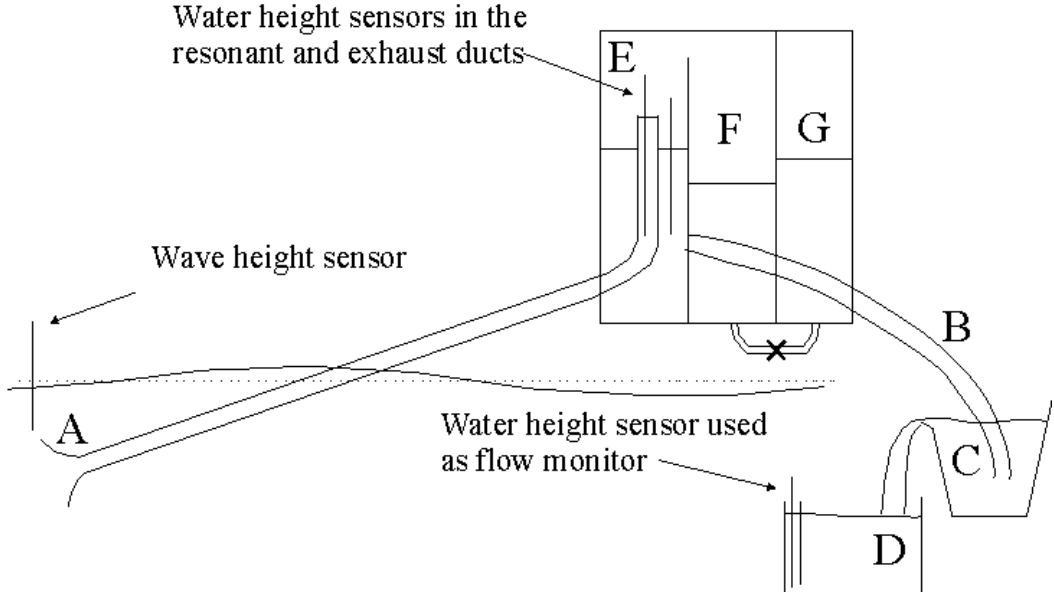,width=1\linewidth}
\end{center}
\caption{Schematic diagram of the scale
model tested in the wave tank (see text).}
\label{figexpsetup}
\end{figure}

In this section we describe wave tank experiments conducted with a scale
model of the pump driven by polychromatic waves. The model was
designed on a 1:20 scale with respect to a full-scale oceanic application and is similar to
that used for the monochromatic wave experiments reported in \cite
{czitrom2000b}. A schematic diagram of the experimental setup is shown in figure
\ref{figexpsetup}.

The mouth of the resonant duct (A) was placed in the wave tank exposed to
the passing waves, while the exhaust duct (B) was placed in a bucket of
water (C), which was leveled flush with the surface of the tank. Water
spilt from the bucket was collected in a recipient (D). The pump was primed
at the start of the experiments by creating a partial vacuum that brought
water up from the tank and bucket, to a working level in the compression
chamber (E). We define the portion of resonant duct extending above this
level as the \textit{sill height}. The volume of air in the compression
chamber was modified by changing the level of water in the connecting
chamber (F); through interchange with the storage tank (G). The model was
fully instrumented with water height sensors of the capacitance type in the
resonant and exhaust ducts, as well as in the water deposit (D), used to
monitor flow through the pump continuously. A water height sensor was also
placed at the wave tank surface, above the resonant duct mouth, to measure
the driving wave height. The sensors were sampled at 0.1s intervals.
Dimensions of the various components are explicited in Table I.


\begin{center}
\begin{tabular}{|l|l|l|}
\hline
$L_1$ & $D_1$ & $A_1$ \\ \hline
5.25m & 0.056m & 0.00985m$^2$ \\ \hline
\end{tabular}
\begin{tabular}{|l|l|l|}
\hline
$L_2$ & $D_2$ & $A_2$ \\ \hline
16m & 0.036m & 0.00407m$^2$ \\ \hline
\end{tabular}

\begin{tabular}{|l|l|}
\hline
$D_{c}$ & $A_{c}$ \\ \hline
0.145m & 0.06605m$^2$ \\ \hline
\end{tabular}
\begin{tabular}{|l|l|}
\hline
$V_0$ & {Sill height} \\ \hline
0.033m$^3$-0.073m$^3$ & 0.005m-0.03m \\ \hline
\end{tabular}

\begin{tabular}{|ll|}
\hline
\multicolumn{1}{|l|}{Wave tank depth} & Resonant duct mouth depth \\ \hline
\multicolumn{1}{|l|}{0.86m} & 0.279m \\ \hline
\end{tabular}

\vspace{0.1in}
{\small Table I. Dimensions of the various components of the model pump.}
\end{center}

\begin{figure}[t]
\begin{center}
  \epsfig{file=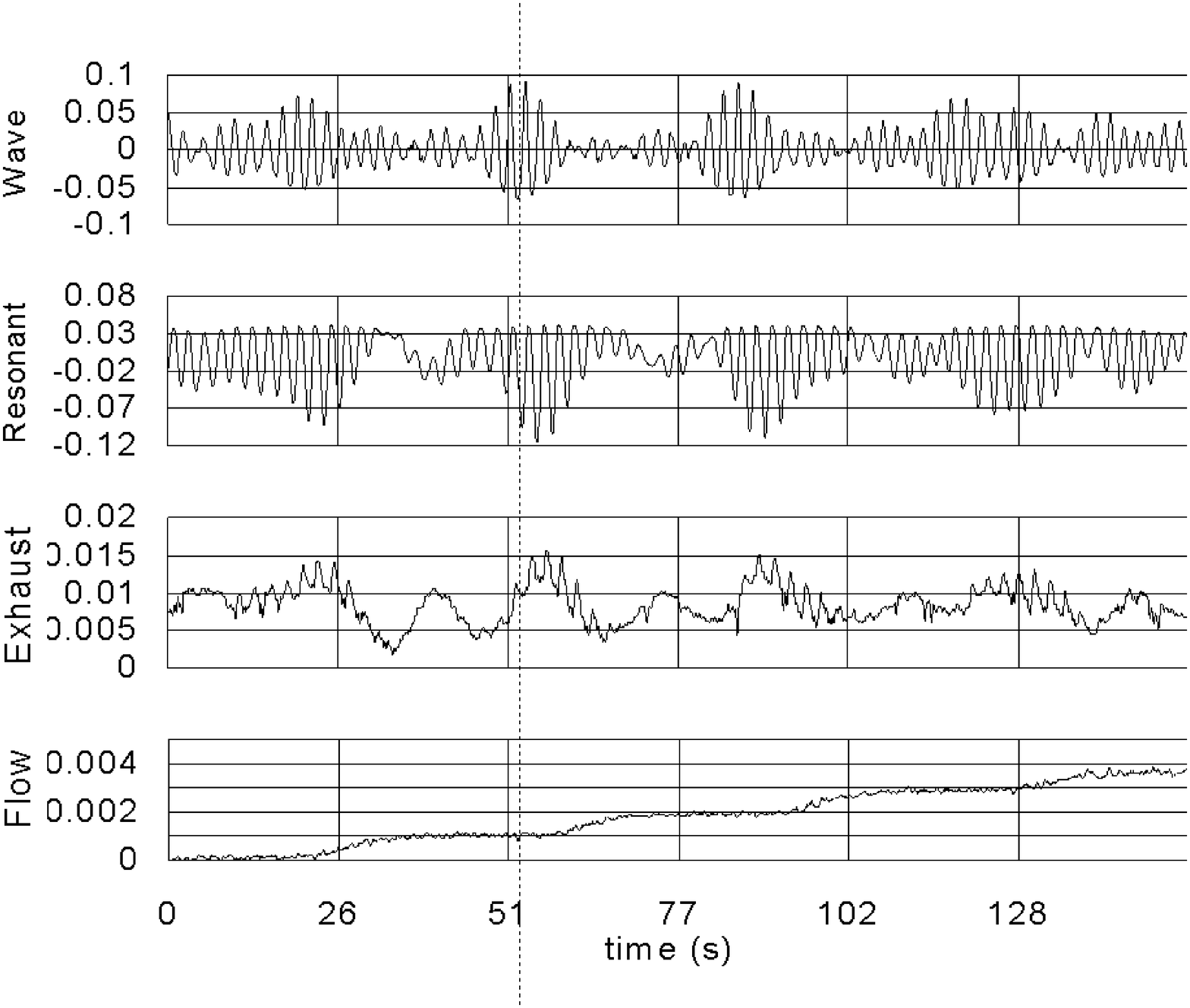,width=1\linewidth}
\end{center}
\caption{Time series of the water height
measuremments in the wave tank, the resonant and exhaust ducts (in metres) and the
flow recipient (in cubic metres).}
\label{figtraces}
\end{figure}

The model pump was driven with waves of various spectra, which were designed
to mimic real sea surface waves and also to allow for differentiation of the
response of the system to each frequency component. For the data shown in this paper, the wave generator was
programmed to include four different wave periods (1.8, 2, 2.2 and 2.4
seconds) while keeping the amplitude at 0.04 m for all components. Traces in
time obtained from the different sensors for one of the experiments are
shown in figure \ref{figtraces}.

It can be seen in this figure that the pulsating forcing wave drives the
response in the resonant duct and in the flow. The regions with largest
slope in the flow trace lag behind the largest oscillations in the resonant
duct which in turn lag behind the driving wave pulsations. From this
phenomenon we can characterize a response time for the experimental system,
between the largest wave oscillations and the largest observed flow rate, of
approximately 12 seconds. It can also be seen that the events of large
oscillations in the resonant duct, which produce pumping, are associated
initially with an increased water level in the exhaust duct side of the
compression chamber, which subsequently oscillates with a period greater
than the driving wave. This lower frequency is related to the second natural
frequency of oscillation of the sea-water pump, identified as $\omega _0^{-}$
in expression (\ref{frec}), and is excited by the change in level which
results from spilling in the compression chamber.

\begin{figure}[t]
\begin{center}
  \epsfig{file=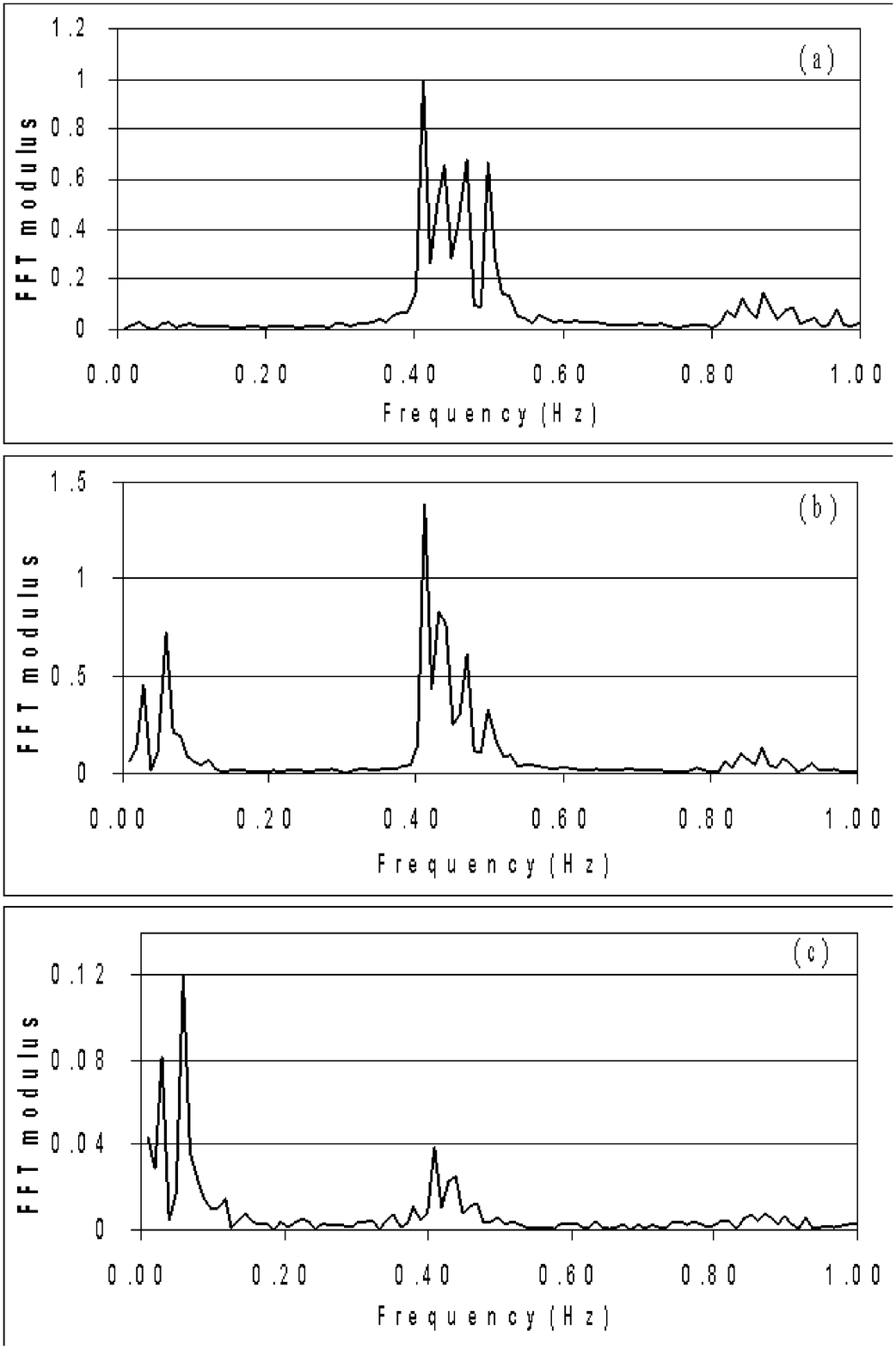,width=0.65\linewidth,bbllx=20,bblly=0,bburx=640,bbury=982,clip=}
\end{center}
\caption{Fast Fourier Transforms of (a) the
driving wave height, water height in the (b) resonant duct and (c) exhaust
side of the compression chamber. These FFTs correspond to the time series shown in figure \ref{figtraces} and are all normalized to the maximum of (a).}
\label{figFFT}
\end{figure}

In figure \ref{figFFT}, Fast Fourier Transforms (FFT) of 1024 data points of
the forcing wave signal and the oscillations within the ducts are shown. Comparing figures \ref{figFFT}.a and \ref{figFFT}.b, it can be seen that, for the resonant duct, the main
driving wave frequencies are transformed so that the resonant frequency peak is
clearly larger than the others for that particular experiment. The
lower frequencies also present in the resonant duct response correspond, on
the one hand, to the second natural frequency of oscillation noted in the
previous paragraph, and to combination frequencies $nf_i-mf_j$ where $n$ and $m$ are integers ($n=m=1$ most important) and $f_i$ and $f_j$ are any of the main driving
frequencies. These combinations are to be expected considering that the
frequency of the pulsations in the driving wave must correspond to
differences between the main forcing frequencies. In the exhaust duct
(figure \ref{figFFT}.c), these low frequency components are the most
important part of the oscillation.

\section{Tuning to Polychromatic Waves}
\label{tuning}

\subsection{FFT analysis}

\begin{figure}[t]
\begin{center}
  \epsfig{file=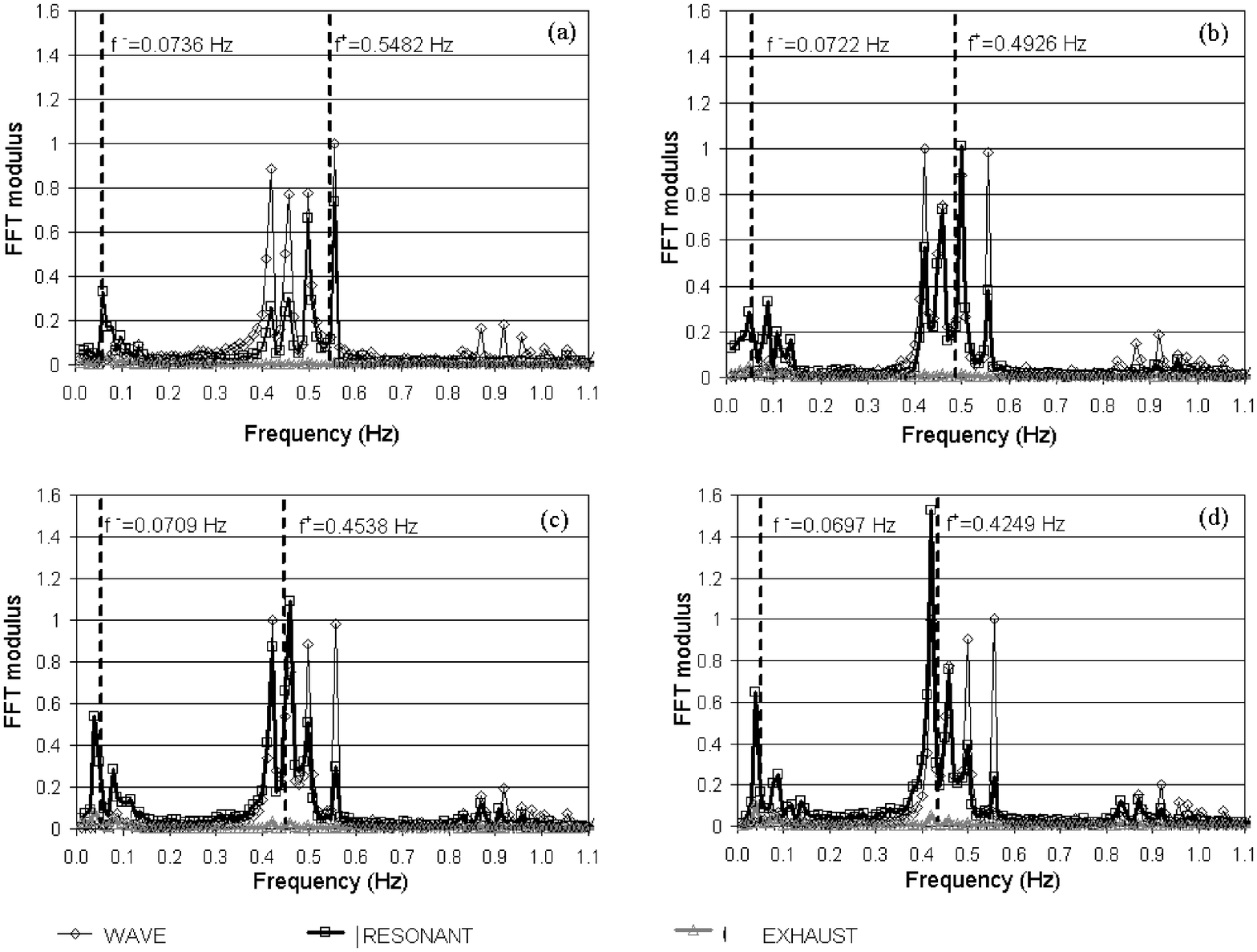,width=1\linewidth}
\end{center}
\caption{Fast Fourier Transforms of
experimental data of the forcing wave ($\Diamond$) and the oscillations in the ducts ($\Box$: resonant duct signal; $+$: exhaust duct signal, very weak with respect to the driving wave and the resonant duct).
The same driving wave of 4 period components $T_{\mathrm{poly}}=1.8, 2, 2.2$ and 2.4 was used in all cases. The volume of air in the compression chamber $V_{0}$ was varied in the four cases shown to match the value determined by each value of $V_{0Lin}(T_{\mathrm{poly}})$ (see text). Its value in m$^3$ was (a) 0.0058, (b) 0.0074, (c) 0.0091 and (d) 0.011. The two natural frequencies of oscillation
for each configuration are shown as vertical lines.}
\label{figFFTtuning}
\end{figure}

In figure \ref{figFFTtuning}, we show FFT's of the forcing wave signal and of the
oscillations in the two ducts measured in four different cases. These four
experiments were conducted in the same wave conditions (a polychromatic wave with four components  $T_{\mathrm{poly}}=1.8, 2, 2.2$ and 2.4) and only the volume of air in the compression chamber had a different value in each case, corresponding to the prediction $V_{0Lin}$ of equation \ref{volin} for each period component.
Also shown in these figures are the two natural frequencies of oscillation ($\mathrm{f}^{+}$ and $\mathrm{f}^{-}$ $=1/T$)
obtained from equation (\ref{frec}) for each volume of air in the compression chamber. The energy in the forcing wave ($\Diamond$'s in figure \ref{figFFTtuning}), which is injected mainly in the four principal peaks of the signal, is redistributed in the response of the system. The frequency content in the resonant duct signal ($\Box$'s in figure \ref{figFFTtuning}) is clearly different for each case and it can be seen that the largest peak occurs near the natural frequency of oscillation $\mathrm{f}^{+}$
predicted by the linearized model. It is apparent that the system responds with
greater strength to the frequency nearest its natural frequency of
oscillation, despite the polychromatic nature of the forcing wave.

\subsection{Tuning from the experimental data}

\begin{figure}[t]
\begin{center}
  \epsfig{file=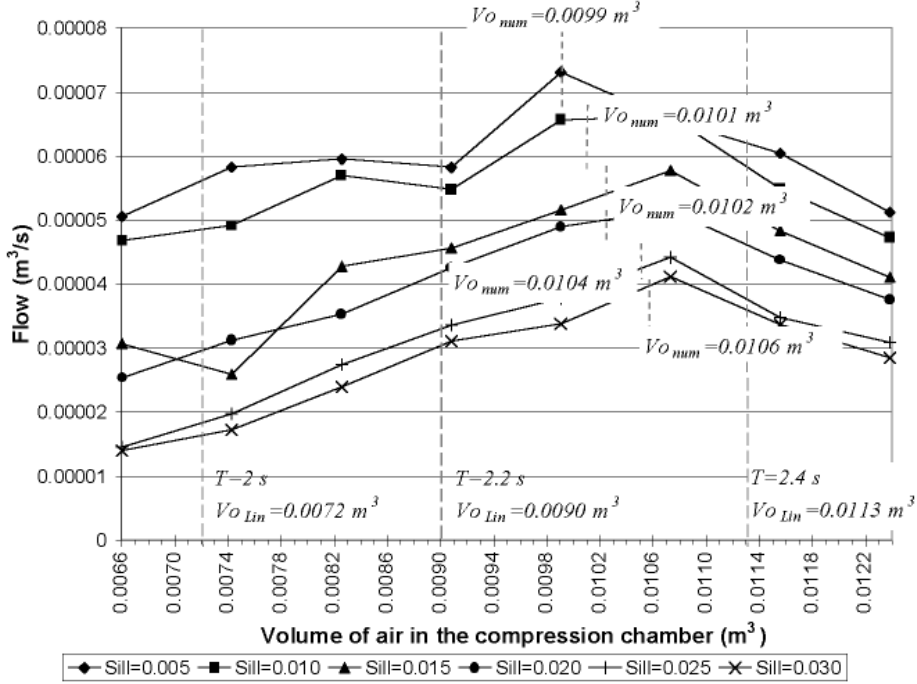,width=1\linewidth}
\end{center}
\caption{Average flow rate obtained in
the experiments vs. volume of air in the compression chamber for the same forcing
wave of figure \ref{figFFTtuning}. The long vertical lines show the linearized model prediction $Vo_{Lin}$
for each component. The short vertical lines show the numerical model prediction $V_0$
for the greater period and each sill height (given in meters).}
\label{figflow}
\end{figure}

The air chamber volume for which flow through the pump is greatest for each particular configuration
(henceforth referred to as the resonant volume), can be
determined from the experimental flow data. Plots of average flow rate as a
function of the volume of air in the compression chamber for various sill
heights are shown in figure \ref{figflow}, for the same 4-component forcing
wave of figure \ref{figFFTtuning}.
The long vertical lines show the linearized model prediction of the resonant air volume ($Vo_{Lin}$) for each
forcing wave frequency component (the prediction for the 1.8 seconds
component lies below the range covered in this plot). The vertical line furthest
to the right corresponds to the largest wave period present (2.4 s). If we
consider the monochromatic responses, according to the algorithm for
resonant flow $(Q_0)$ mentioned above and given that all four main components had essentially the same amplitudes, we should expect the response to the
component of the largest period to produce the greatest pumped flow. In figure \ref{figflow}, the observed maximum flow rates
occur for volumes of air in the compression chamber bounded by the linearized model predictions ($Vo_{Lin}$) for the two largest
periods. This suggests that the components of larger periods determine the collective resonant response of the system to polychromatic waves. It can also be seen in figure \ref{figflow}, that the resonant volume, as
defined by the maximum average flow rate is, for all sill heights, always
smaller than the $Vo_{Lin}$ for the maximum period component. This is
consistent with the monochromatic experiments of \cite{czitrom2000b} who observed that pumping shifts
the actual resonant volume to smaller values than the linearized model
prediction. Similar to the monochromatic case,
greater pumping induces a greater shift towards smaller resonant air chamber
volumes. The short vertical lines in figure \ref{figflow} are the values for the resonant volume
obtained from monochromatic numerical model experiments with the greatest
period and six sill heights. These were obtained by running the model for a
series of compression chamber air volumes covering the range shown in figure
\ref{figflow}. It is clear that the resonant air chamber volume predicted by
the monochromatic model closely resembles the observed volume that produces maximum for
each case. This result encourages us to build a tuning criterion for waves
with various frequency components based on the numerical monochromatic
algorithms for $V_0$ and $Q_0$.

It should be noted that a forcing wave with the same amplitude for all of
its' components is, of course, unlikely a common real sea-surface wave.
Thus, the greatest pumping can occur when the system is tuned to a component
that is not necessarily the one with the largest period. In the next section
we use the flow prediction obtained numerically for monochromatic waves ($%
Q_0 $), which takes into account the wave amplitude for a given frequency,
and use it to choose an optimal tuning condition.

\subsection{Numerical experiments with polychromatic waves}

\begin{figure}[t]
\begin{center}
  \epsfig{file=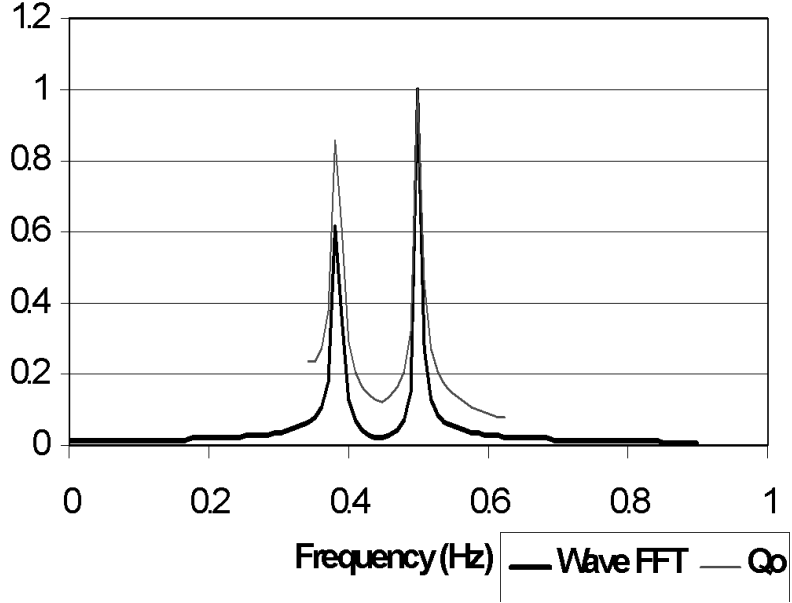,width=0.75\linewidth,bbllx=20,bblly=0,bburx=1020,bbury=780,clip=}
\end{center}
\caption{Fast Fourier Transform of the
forcing wave signal and the corresponding graphic for $Q_0(T)$ for a two-component forcing wave (of 2 and 2.6 seconds periods). Each curve
is normalized with respect to its maximum.}
\label{figFlowvsFFT}
\end{figure}

We performed a series of numerical simulations using the same configuration of the wave tank
experiments, but driving the model with different synthetic waves of multiple frequency components.
The mathematical expression of the forcing waves used in the numerical experiments was:

\begin{equation}\label{forcing_form}
    \sum_{i}A_i\sin\frac{2\pi t}{T_i}
\end{equation}

We report on two particular cases that let us discuss the basic features of the polychromatic response: In the first case, the forcing signal was a two-component wave (of $T_1=2$ and $T_2=2.6$ seconds)
whose FFT is shown in a heavy line in figure \ref{figFlowvsFFT}. We used a greater
amplitude for the higher frequency deliberately $A_2=0.6A_1$, to show that the wave
amplitude must also be used in determining the maximum flow through the
pump. The light line in the same figure is the monochromatic flow prediction $Q_0$, calculated with the frequency $f=T^{-1}$ and
corresponding amplitude values from the FFT. It can be seen that, in
contrast with the experimental observations which were conducted with the
same amplitude for all frequency components, here we find that the smaller
period produces a greater flow. Clearly, the amplitude of each component
must be taken into account to choose the most efficient tuning condition. We remark, however, that the ratio between the expected
maximum flows ($Q_0$ in figure \ref{figFlowvsFFT}) for the two frequencies, is noticeably closer to 1
than the ratio between the amplitudes of each component in the FFT of the
forcing wave. This shows that energy from components with larger periods is
used more efficiently by the system.

\begin{figure}[t]
\begin{center}
  \epsfig{file=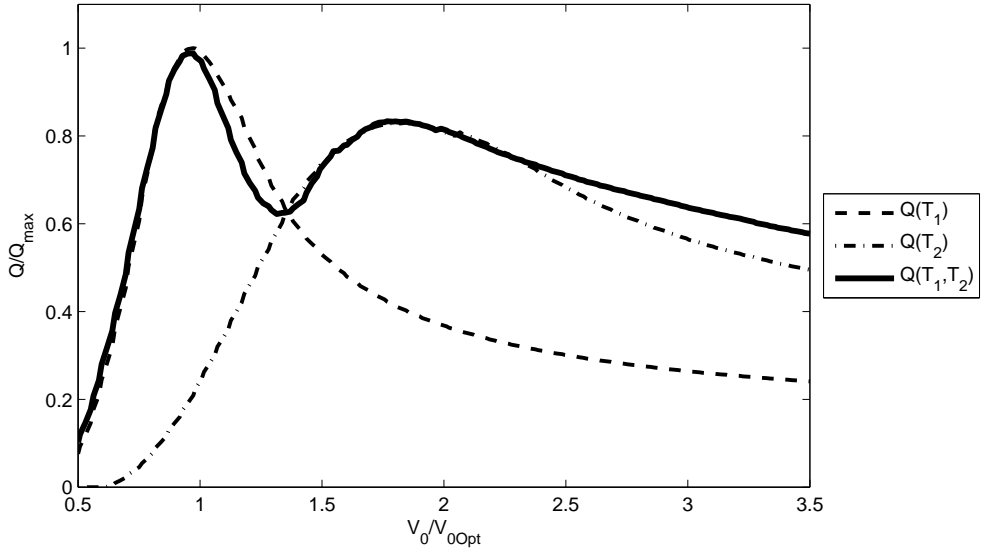,width=1\linewidth}
\end{center}
\caption{Flow vs. volume of air in the
compression chamber obtained with the numerical model forced by: a
signal with components of $2$ and $2.6\;s$ ($Q(T_1,T_2)$); only the $2$s component ($Q(T_1)$);
only the $2.6$s component ($Q(T_2)$).}
\label{figFlowvsVo}
\end{figure}

Once we have chosen the wave frequency component which produces the greatest
flow, the appropriate air chamber volume can be calculated using the
monochromatic numerical algorithm for $V_0$. In order to test the effectiveness of this criterion as a selection
tool for the optimal volume of air of the compression chamber, in figure \ref{figFlowvsVo} we show the flow rate through the pump plotted against volume of air in the compression chamber, computed with the numerical model forced by the wave signal shown in figure \ref{figFlowvsFFT} ($Q(T_1,T_2)$).  Also shown in this figure ($Q(T_1)$ and $Q(T_2)$) are the
flow rates obtained by forcing the numerical model with monochromatic waves
of frequencies and amplitudes corresponding to each one of the components in
figure \ref{figFlowvsFFT}. The abscissa in figure \ref{figFlowvsVo} is normalized with
the optimal volume of air predicted by $V_0$ for the higher frequency. The
flow response for the polychromatic forcing ($Q(T_1,T_2)$) has two maximums, which
correspond to the resonant volumes for each frequency component. It is
confirmed that the absolute maximum corresponds to the one predicted by $Q_0$
in figure \ref{figFlowvsFFT} (1 in the horizontal scale). The correspondence of the maximum flows in the poly and monochromatic wave
forcings in figure \ref{figFlowvsVo} shows that the attainable flow with one
frequency is essentially not affected by the presence of the other. That is, for this separation between frequencies, the pump behaves similar to a flute which amplifies each frequency distinctly, depending on the natural frequency of oscillation to which it is tuned.

\begin{figure}[t]
\begin{center}
  \epsfig{file=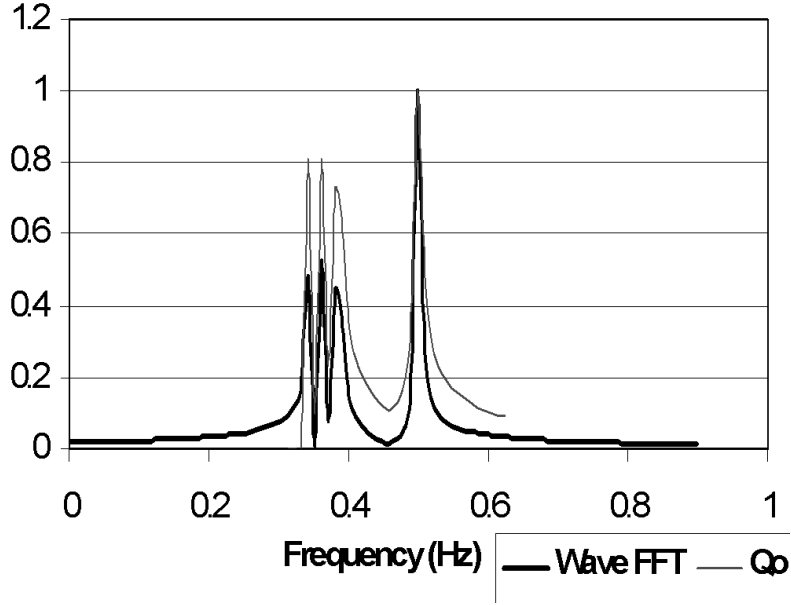,width=0.75\linewidth,bbllx=20,bblly=0,bburx=1020,bbury=780,clip=}
\end{center}
\caption{Fast Fourier Transform of the
forcing wave signal and the corresponding graphic for $Q_0(T)$ for a four-component wave (of 2.9, 2.75 and 2.6 and 2 seconds). Each curve
is normalized with respect to its maximum.}
\label{figFlowvsFFT2}
\end{figure}

\begin{figure}[t]
\begin{center}
  \epsfig{file=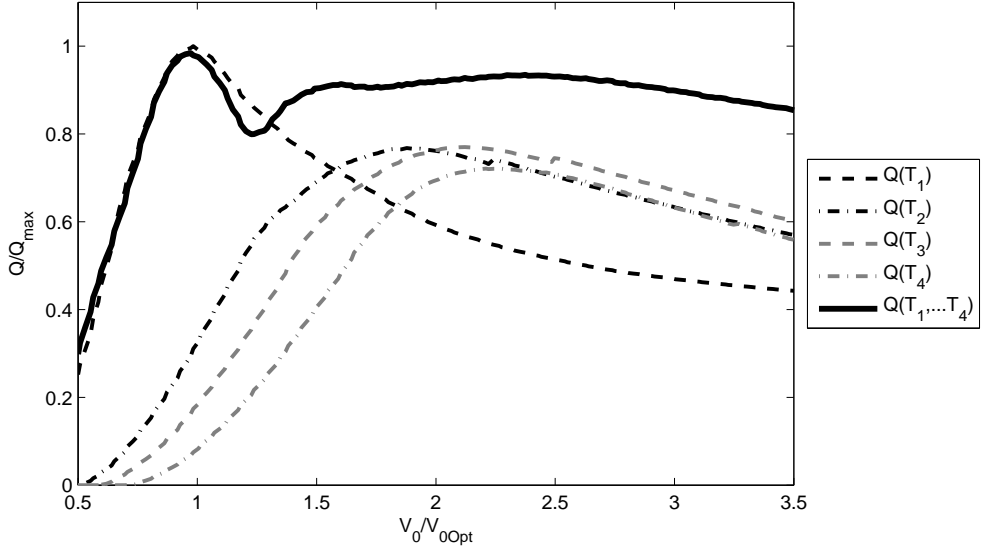,width=1\linewidth}
\end{center}
\caption{Flow vs. volume of air in the
compression chamber obtained with the numerical model forced by: a signal
with components of $T_i=2,2.6,2.75$ and 2.9 seconds (solid line, $Q(T_1,\ldots T_4)$) and four series obtained with monochromatic forcings for each component ($Q(T_i)$).}
\label{figFlowvsVo2}
\end{figure}

We now analyze another case in which the forcing wave has four components of periods $T_i=2,2.6,2.75$ and 2.9 seconds.
As it can be seen in figure \ref{figFlowvsFFT2}, the signal is composed of a \textit{block} of three frequencies close to each other (corresponding to the wave periods of 2.9, 2.75 and 2.6 seconds) plus a fourth frequency, slightly more
separated, corresponding to $T_1=2$s. As with the previous case, we attempt using the $Q_0$ curve to determine
which component we should choose for tuning the system. If we guide
ourselves only by the maximum flow value, we would again choose the 2
seconds component that is separated from the block. Similar to figure \ref
{figFlowvsVo}, we used the numerical model to test the validity of this tuning
criterion as shown in figure \ref{figFlowvsVo2}. Here it can be seen that, while
the absolute maximum flow rate in the polychromatic response ($Q(T_1,\ldots T_4)$) is
correctly identified with the corresponding monochromatic response ($Q(T_1)$), the
flow values $Q(T_1,\ldots T_4)$ for the air volumes resonant with the lower frequencies are clearly higher than the
maxima of the monochromatic responses ($Q(T_2)$, $Q(T_3)$ and $Q(T_4)$). This shows that
the responses to the three lower frequencies in the forcing wave interact to give a larger
flow than expected from each component separately and indicates the existence of a {\it bandwidth} of resonance. The bandwidth in oscillating systems is related to damping \cite[see e.g.][]{falnes2002}. For the seawater pump damping is dominated by the pumping process itself : when the amplitude of oscillation increases in the resonant duct, water is spilt at the sill during a larger part of the period. This determines that, in the pumping regime, the limit imposed on the amplitude by the pumping process dominates over other damping mechanisms such as friction or vortex formation. The resonance bandwidth of the pumping system can be defined using different norms. We choose it to be the period interval for which flow through the pump is at least 80\% its maximum value, and estimate it using flow data (not shown here) from the scale model experiments with different monochromatic forcing periods. For the experimental configuration of figure \ref{figFlowvsVo2}, a resonance bandwidth covering a broad period interval between 1.5 and 3 seconds is obtained. This rather large bandwidth can explain the previously observed fact that the polychromatic response $Q(T_1,\ldots T_4)$ in figure \ref{figFlowvsVo2} shows greater pumped flow, at the air volumes resonant with the lower frequencies, than any of the monochromatic responses ($Q(T_2)$, $Q(T_3)$ and $Q(T_4)$).

\begin{figure}[t]
\begin{center}
  \epsfig{file=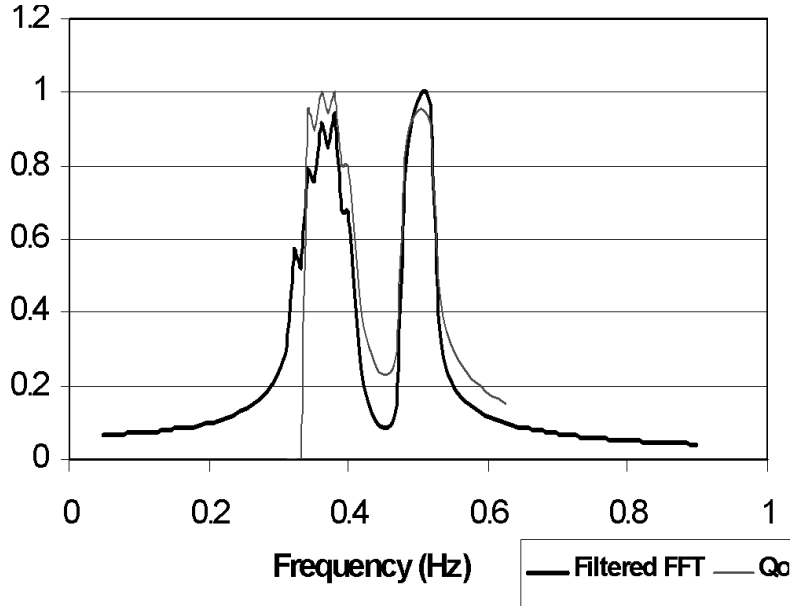,width=0.75\linewidth,bbllx=20,bblly=0,bburx=1020,bbury=780,clip=}
\end{center}
\caption{Filtered FFT of the forcing
wave signal of figure \ref{figFlowvsFFT2} with a running mean of 0.04 Hz and the corresponding $Q_0(T)$
curve; each curve is normalized with respect to its maximum.}
\label{figFlowvsFFT2filter}
\end{figure}

In real sea-surface waves, some or all of the frequencies present in a
block are more likely to persist in time, than a single frequency on its'
own. Furthermore, because of the resonance bandwidth discussed in the previous paragraph, a better pump performance can be expected when exploiting the collective response of the system to various peaks of a block. An ultimate tuning criterion for irregular sea waves should therefore take into account information on the frequency-domain distribution of the forcing. We propose using a running mean of a given FFT as a
method of pondering the relative importance of blocks of frequencies
versus single frequencies. Such a running mean would diminish the amplitude
of single peaks, while allowing blocks to better retain their
relative weight. As an example of how this procedure could work, in
figure \ref{figFlowvsFFT2filter} a running mean of the FFT in
figure \ref{figFlowvsFFT2} with a width of 0.04 Hz is shown (we choose the width for the filtering as 10\% of the estimated bandwidth, which was $\approx 0.4$Hz for this case). As expected, the reduction of the single
frequency peak is noticeably greater than that of the three frequency
block. Also shown in this figure is the expected flow $Q_0$ as calculated
using the filtered FFT amplitudes. It is clear that a somewhat greater flow
is now associated to the block of frequencies than to the single frequency
peak.

Using this type of filtering of the frequency content of the forcing wave signal can be an important tool in the design of the tuning algorithm for the sea-water pump because, as mentioned before, in a constantly changing sea, blocks of frequencies in the forcing spectrum have the added convenience of a likely greater persistence in time than single frequency peaks.

\section{Conclusions}
\label{concl}

The response of an oscillating-water-column (OWC) seawater pump to polychromatic waves was studied using wave tank tests with a scale model and numerical simulations. We show that, in spite of being driven by
polychromatic waves, the system responds mainly in a monochromatic way, amplifying a selected frequency from the wide spectrum present in its
forcing. The measured flow through the pump as a function of the experimental tuning parameter (the volume of air in the compression chamber) retrieves thus the same basic qualitative features that were reported by \cite{czitrom2000b} for experiments with monochromatic waves. The main observations are that the natural frequency of oscillation of the OWC decreases due to the overtopping associated with the pumping process, and that greater pumping is obtained when tuning the system to wave components of lower frequencies, given equal amplitudes. For the experiments reported here, the observed \emph{resonant volume}, defined as the volume of air in the compression chamber for which flow through the pump is greatest, is predicted remarkably accurately by monochromatic algorithms using the lowest frequency peak present in the forcing spectrum. Numerical experiments with synthetic forcing waves showed nonetheless that the polychromatic response of the system is not always merely a superposition of monochromatic responses. A cooperative interaction between the responses to slightly different frequencies present in the forcing wave was evidenced, confirming the existence of a resonance bandwidth and hinting to the use of a frequency-space-averaged signal as input of the proposed polychromatic tuning algorithm. The latter is based on the monochromatic algorithms for resonant volume
of air in the compression chamber ($V_0$) and resonant flow rate ($Q_0$).

\begin{figure}[t]
\begin{center}
  \epsfig{file=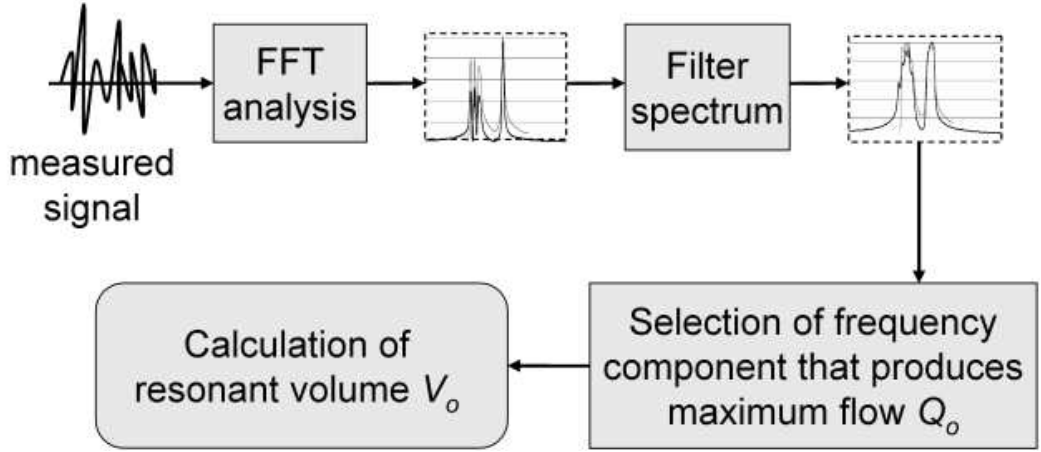,width=1\linewidth}
\end{center}
\caption{Tuning protocol diagram (see text for details).}
\label{algoritmo}
\end{figure}

The tuning process proposed for the sea-water pump in an irregular sea
consists of various steps that can be easily performed by a pre-programmed microchip and is summarized in figure \ref{algoritmo}. First, the FFT analysis would be performed
on a series of measurements taken with a wave sensor, and the resulting spectrum would be filtered to
ponder the frequency content of the signal. The maximum expected flow would then be identified using the $Q_0$ calculation
from the filtered amplitudes and corresponding frequencies and, finally, the resonant volume of air for optimal sea-water pump
performance would be computed using the $V_0$ algorithm with the
appropriate frequency and unfiltered amplitude of the chosen component. This tuning procedure thus outputs the volume of air in the compression chamber that will give the best pump performance. An actual control system modifying the volume of air in the compression chamber could then perform its task at any given time taking only as input a series of measurements of the wave signal at the resonant duct mouth.

\vspace{0.6cm}

\noindent {\bf \textbf{Acknowledgments}}

\vspace{0.2cm}

We wish to thank Tony Dalrymple, Joe Hammack, Nobu Kobayashi and
Brad Johnson for allowing and helping us to use the wave tanks at the Center
for Applied Coastal Research, The University of Delaware. Thanks are also
due to Ranulfo Rodr\'\i guez Sobreyra for developing and implementing the
sensing and data acquisition system used during the experimental work, and
Esteban Prado, Catalina Stern and Arturo Olvera for useful discussions.
We greatly acknowledge the John D. and Catherine T. MacArthur Foundation for
their generous support through
a grant from the Fund for Leadership Development. Support was also provided
by the Direcci\'on General de Apoyo al Personal Acad\'emico (Projects
IN-104193, 106694, 107197 and a Graduate Studies Scholarship for RGD) and
the Instituto de Ciencias del Mar y Limnolog\'\i a (Project ''Energ\'\i a de
Oleaje'' No. 139), both from the National University of Mexico (UNAM).
Support was also provided by CONACyT through project 'Matem\'aticas
no-lineales en la f\'\i sica y en la ingenier\'\i a' (FENOMEC), No.
625427-E.

\end{document}